\documentclass[%
reprint,
 amsmath,amssymb,
prl
]{revtex4-2}

\usepackage{graphicx}
\usepackage{dcolumn}

\usepackage[labelfont=bf]{caption}
\usepackage{bm}
\usepackage[english]{babel}
\usepackage[utf8]{inputenc}
\DeclareUnicodeCharacter{2212}{-} 
\DeclareUnicodeCharacter{03B4}{\ensuremath{\delta}} 
\usepackage{siunitx} 
\usepackage{textcomp} 
\usepackage[version=4]{mhchem}
\DeclareSIUnit\angstrom{\text{\AA}} 
\usepackage{enumitem} 

\usepackage[table,xcdraw,svgnames]{xcolor} 
\usepackage[pdfencoding=auto,psdextra,colorlinks]{hyperref} 
\AtBeginDocument{\hypersetup{ 
  colorlinks   = true, 
  urlcolor     = purple, 
  linkcolor    = teal, 
  citecolor   = violet 
}}

\usepackage{xcolor, soul}
\sethlcolor{yellow}

\usepackage{subcaption} 
\usepackage[most]{tcolorbox}
\usepackage{xcolor}

\usepackage{tabularx} 
\usepackage{booktabs} 
\usepackage{multirow} 

\usepackage{duckuments}

\begin{document}

\tcbset{
  graybox/.style={
    colback=gray!20,   
    colframe=gray!20,  
    boxrule=0pt,       
    arc=1mm,           
    left=2mm, right=2mm, top=1mm, bottom=1mm, 
  }
}

\tcbset{
  orangebox/.style={
    colback=orange!20,   
    colframe=orange!20,  
    boxrule=0pt,       
    arc=1mm,           
    left=2mm, right=2mm, top=1mm, bottom=1mm, 
  }
}

\title{Polar discontinuity screening by charge disproportionation in ferroelectric-nickelate superlattices}

\author{Edith Simmen$^1$}
    \email{edith.simmen@mat.ethz.ch}
\author{Jonathan Amara$^1$}
\author{Philippe Ghosez$^2$}
\author{Nicola A. Spaldin$^1$}%
\affiliation{%
$^1$ Materials Theory, ETH Zurich, 8093 Zurich, Switzerland, 
$^2$ Theoretical Materials Physics, Q-MAT, Université de Liège, B-4000 Sart-Tilman, Belgium
}%

\date{\today}

\begin{abstract}
We use density functional theory (DFT) to demonstrate rearrangement of charge disproportionation as a new mechanism for screening polar discontinuities at heterointerfaces. We focus on SmNiO$_3$, a III-III perovskite characterized by an insulating ground state that originates from disproportionation of the electrons at the Ni sites. When interfaced with ferroelectric, II-IV perovskite BaTiO$_3$, this system can display a polar discontinuity whose magnitude is determined by the nature of the interface (NiO$_2$–BaO or SmO–TiO$_2$) and the orientation of the BaTiO$_3$ spontaneous polarization. In the orientation maximizing the interface charge, SmNiO$_3$ compensates by transferring \SI{1}{\elementarycharge\per\text{f.u.}} between the two interfaces and modifying the disproportionation sequence from Ni$^{4+}$/Ni$^{2+}$ to Ni$^{2+}$/Ni$^{2+}$ at the SmO–TiO$_2$ interface and to Ni$^{4+}$/Ni$^{4+}$ at the NiO$_2$–BaO interface. While this mechanism is potentially universal in all perovskites exhibiting charge disproportionation, our formal analysis of BiNiO$_3$/BaTiO$_3$ superlattices suggests that larger polar discontinuities may need more complex disproportionation patterns for screening.  

\end{abstract}

\maketitle

\paragraph{Introduction–} 
Understanding the physics of interfaces in thin films is vital for the development of next-generation electronic devices, as these boundaries can fundamentally alter film properties and lead to phenomena absent in bulk counterparts. In perovskite oxide heterostructures, such interface-driven effects include the suppression of the spontaneous polarization in thin-film ferroelectrics \cite{mehta_depolarization_1973, junquera_critical_2003, efe_engineering_2024}, the two-dimensional electron gas (2DEG) at the interface of LaAlO$_3$ and SrTiO$_3$ \cite{ohtomo_highmobility_2004, nakagawa_why_2006, bristowe_origin_2014, lemal_polarityfield_2020}, and the emergence of ferroelectric topological textures in PbTiO$_3$/SrTiO$_3$ superlattices \cite{yadav_observation_2016,das_observation_2019,junquera_topological_2023}. Each of these phenomena stems from uncompensated interface charges, either from the ferroelectric polarization, a built-in polar discontinuity, or both. 

Such a built-in polar discontinuity arises at the interface between two perovskite oxides with differently charged sublayers. We refer to the contribution of these charged layers to the overall bulk polarization as the `layer polarization' \cite{spaldin_layer_2021, efe_happiness_2021, simmen_interplay_2025}; alternatively, this quantity is also known as the `formal polarization' \cite{stengel_berryphase_2009, bristowe_origin_2014, fang_how_2023}. II-IV perovskites such as BaTiO$_3$ or SrTiO$_3$ have formally uncharged AO and BO$_2$ sublayers and their centrosymmetric unit cell contains zero layer polarization. 
In III-III perovskites such as LaAlO$_3$ or SmNiO$_3$, the sublayers (AO)$^+$ and (BO$_2)^-$ are oppositely charged and even the centrosymmetric structure has a non-zero layer polarization \cite{stengel_electrostatic_2011, spaldin_layer_2021}. Perovskite oxides with this layer polarization have a charged (001) surface and a polar discontinuity when interfaced with a II-IV perovskite. 

The polar discontinuity leads to unfavorable interface charges that require screening \cite{goniakowski_polarity_2008, hong_screening_2016}. First, screening by free charge carriers can lead to the formation of a 2DEG as occurs at the LaAlO$_3$/SrTiO$_3$ interface \cite{ohtomo_highmobility_2004, nakagawa_why_2006, bristowe_origin_2014, lemal_polarityfield_2020}. These carriers can originate from band bending and Zener breakdown or defects such as oxygen vacancies \cite{bristowe_origin_2014, lemal_polarityfield_2020}. Second, charged defects can provide charge compensation \cite{efe_polarizing_2026, bristowe_origin_2014, zhong_polarityinduced_2010, herranz_high_2007}. Lastly, polar distortions provide another important charge compensation mechanism, particularly at interfaces with ferroelectrics, where the spontaneous polarization orients to reduce the polar discontinuity \cite{murray_theoretical_2009, spaldin_layer_2021, simmen_interplay_2025}.  

Here, we introduce a new mechanism for screening polar discontinuities: a change in the charge-ordering pattern of systems with charge disproportionation. We demonstrate the mechanism for SmNiO$_3$ interfaced with ferroelectric BaTiO$_3$. SmNiO$_3$ is a member of the rare-earth nickelate family and undergoes a characteristic transition from a low-temperature insulating to a high-temperature metallic phase \cite{medarde_structural_1997, catalano_rareearth_2018}. The insulating phase contains charge disproportionated Ni sites \footnote{
Note that the ongoing debate in rare-earth nickelate literature on the nature of the disproportionation \cite{varignon_complete_2017, catalano_rareearth_2018} is not relevant in the context of this work. While the disproportionation was originally described in a completely ionic picture (Ni$^{+2}$ and Ni$^{4+}$) \cite{mizokawa_spin_2000}, the strong hybridization between Ni and O was taken into account more recently, attributing the disproportionation to the formation of a ligand-hole on one of the Ni-O octahedra  ($d^8$ and $d^8\underline{L}$) \cite{johnston_charge_2014}. However, covalency does not change the layer polarization \cite{bristowe_oxide_2009, bristowe_net_2011, bristowe_origin_2014};  per Gauss's law, the charge of an enclosed volume (the NiO$_2$ or SmO sublayers), is independent of its internal distribution. Consequently, using effective Ni charges is sufficient to determine the layer polarization. }, 
formally described as 2~Ni$^{3+}$ $\rightarrow$ Ni$^{2+}$ + Ni$^{4+}$ \cite{mazin_charge_2007, varignon_complete_2017}, in a G-type pattern \cite{alonso_roomtemperature_2000}, leading to alternating formal layer charges of $+1$ (SmO) and $-1$ (NiO$_2$). The interface with BaTiO$_3$ has a polar discontinuity that can be modified in magnitude by the spontaneous polarization of BaTiO$_3$. Our main finding is that the nickelate sublayers adjacent to the interface change their disproportionation pattern to screen the interface charges when the polar discontinuity is sufficiently high. 

\paragraph{Polar discontinuity between SmNiO$_3$ and BaTiO$_3$–} 
First, we find a way to identify the formal charge and the layer charges in bulk SmNiO$_3$. We relax a 20-atom SmNiO$_3$ unit cell in its low-temperature $P2_1/c$ structure and with $A$-type antiferromagnetic ordering (see supplementary material for more details). We strain its in-plane lattice parameters to the  pseudocubic lattice parameter of SrTiO$_3$ (a = \SI{3.898}{\AA} \cite{wahl_srtio3_2008}) to simulate coherent growth on a substrate and use DFT$+U$ as implemented in the VASP code \cite{kresse_efficiency_1996, kresse_efficient_1996} with the PBEsol functional \cite{perdew_restoring_2008} for our calculations (see supplemental material for detailed methods). Due the strong covalent bonding in SmNiO$_3$, simple integration of the charge density around an ionic sites does not lead to a significant difference between the Ni sites \cite{varignon_complete_2017} (Note, however, that covalency does not affect the validity of the polar discontinuity picture \cite{Note1}). 

\begin{figure}
    \centering
    \includegraphics[width=\linewidth]{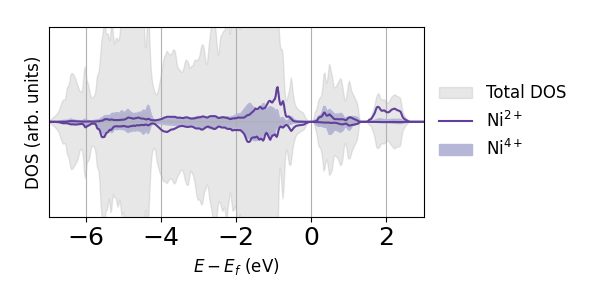}
    \caption{Calculated total DOS of bulk SmNiO$_3$ (grey), with projections on the Ni-$d$ states of one Ni$^{2+}$ (dark purple line) and one Ni$^{4+}$ site (light purple shading).} 
    \label{fig:dos_bulk}
\end{figure}

We therefore identify the charge disproportionation from the density of states (DOS) or the magnetic moments of SmNiO$_3$. Projections of the bulk SmNiO$_3$ DOS onto two of the four Ni sites (purple in Fig.~\ref{fig:dos_bulk}, majority down spin) show a clear difference in the occupancy of the $e_g$ orbitals. The Ni site with +2 effective charge (Ni$^{2+}$, $(t_{2g}^\uparrow)^3(t_{2g}^\downarrow)^3(e_g^\downarrow)^2$) has unoccupied $e_g$ orbitals only for the minority spin (purple line), best visible around \SI{2}{eV}. The Ni site with +4 effective charge (Ni$^{4+}$,  $(t_{2g}^\uparrow)^3(t_{2g}^\downarrow)^3$) has unoccupied $e_g$ orbitals for both spin channels around \SI{1}{eV} (purple area). The same difference is visible in the magnetic moments on the Ni and we use them as a measure of the effective Ni charge. We obtain magnetic moments with absolute values of $\SI{1.3}{\mu_B}$ for Ni$^{2+}$ and $\SI{0.2}{\mu_B}$ for Ni$^{4+}$ in bulk SmNiO$_3$, which agree well with literature \cite{hampel_interplay_2017, varignon_complete_2017}.

\begin{figure}
    \centering
    \includegraphics[width=\linewidth]{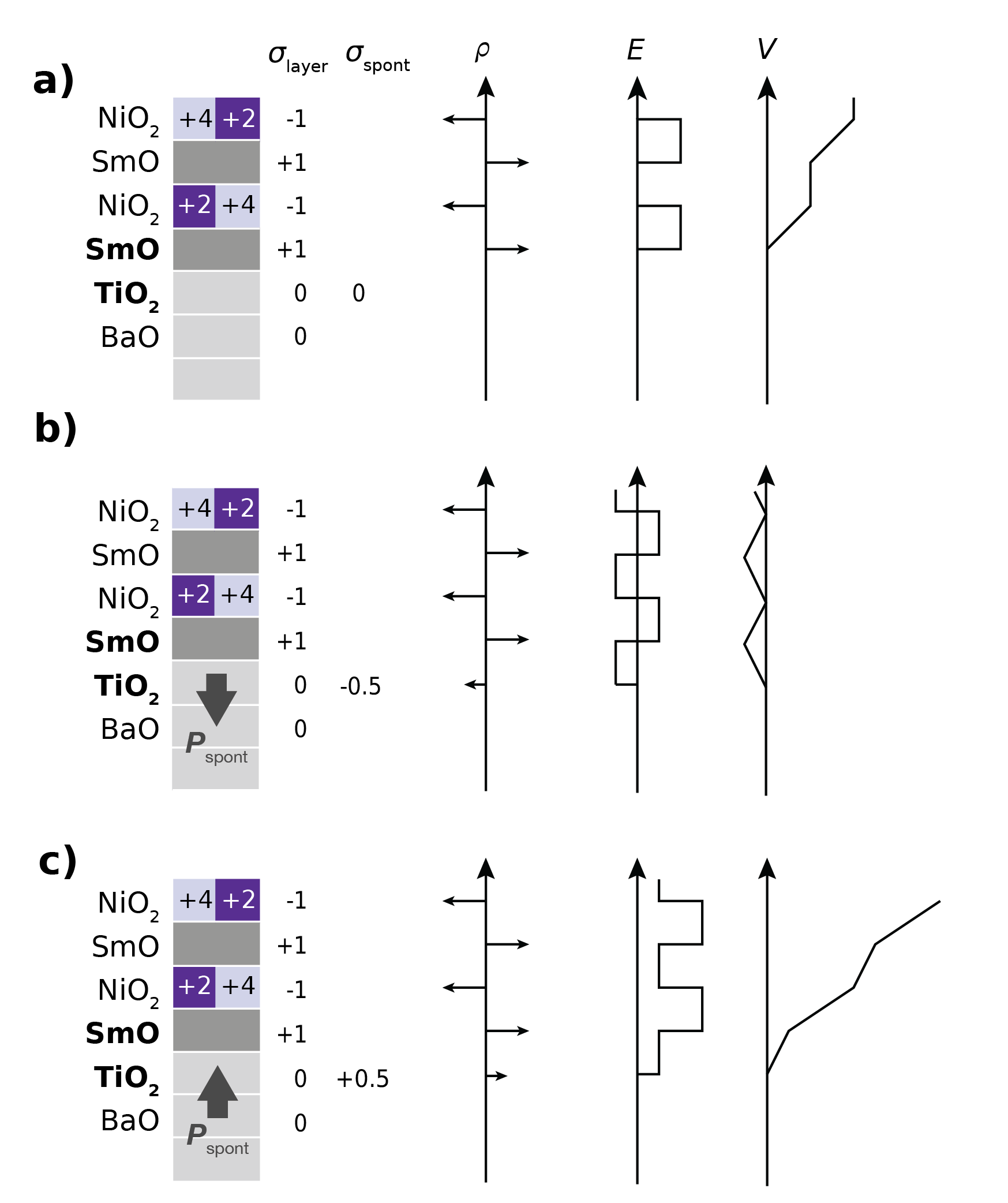}
    \caption{Polar discontinuity at the SmO–TiO$_2$ interface in SmNiO$_3$/BaTiO$_3$ heterostructures. The SmNiO$_3$ and BaTiO$_3$ layer charges ($\sigma_\text{layer}$) and the surface charge from the BaTiO$_3$ spontaneous polarization  ($\sigma_{\text{spont}}$) determine the layer charge density ($\rho$), local electric field ($E$) and the electron potential ($V$) of the SmNiO$_3$ slab. a) With non-polar BaTiO$_3$ ($\sigma_\text{spont} = 0$), the potential inside SmNiO$_3$ diverges. b) Happy orientation: The spontaneous polarization points away from the interface ($\sigma_\text{spont} = -0.5$) and compensates the polar discontinuity. c) Unhappy orientation: The spontaneous polarization points towards the interface ($\sigma_\text{spont} = +0.5$), increasing the polar discontinuity and the divergence of $V$.}
    \label{fig:hp_vs_uhp}
\end{figure}

The interface charge $\sigma_\text{inter}$ originating from the polar discontinuity between SmNiO$_3$ and BaTiO$_3$ along [001] is given by 
\begin{equation}
    \sigma_\text{inter} = \left( \mathbf{P}_\text{BaTiO$_3$} - \mathbf{P}_\text{SmNiO$_3$} \right) \cdot \hat{\mathbf{n}},
\end{equation}
where $\mathbf{P}_\text{SmNiO$_3$}$ and $\mathbf{P}_\text{BaTiO$_3$}$ are the respective bulk polarizations and $\hat{\mathbf{n}}$ is the surface normal \cite{vanderbilt_electric_1993a}. Since the total bulk polarization of centrosymmetric BaTiO$_3$ is zero, the magnitude of the polar discontinuity corresponds to the SmNiO$_3$ layer polarization of $\sim\SI{50}{} \pm n P_\text{q}$. The value of the polarization quantum $P_q$ depends on the choice of unit cell and is $P_\text{q} = \frac{e R_c}{V} = \SI{100}{\micro\coulomb\per\centi\meter\squared}$ for the conventional 5-atom perovskite unit cell with $c$-lattice vector $R_c = \SI{4}{\AA}$ and volume $V$. This results in $\sigma_\text{inter} = \SI{-50}{\micro\coulomb\per\centi\meter\squared}$ (or $\SI{-0.5}{\elementarycharge\per\text{f.u.}}$) for the flat NiO$_2$–BaO and $+\SI{50}{\micro\coulomb\per\centi\meter\squared}$ for the flat SmO–TiO$_2$ interface. The effect of this unscreened interface charge for the SmO–TiO$_2$ interface is illustrated in Fig.~\ref{fig:hp_vs_uhp}a: The charged planes $\sigma_\text{layer}$ lead to an alternating charge density~$\rho$, non-zero electric fields~$E$ and a diverging potential~$V$ inside the SmNiO$_3$ slab. We refer to Ref. \cite{bristowe_origin_2014} or \cite{spaldin_layer_2021} for a more detailed discussion.  

In SmNiO$_3$/BaTiO$_3$ superlattices, the spontaneous polarization of BaTiO$_3$ additionally contributes to $\mathbf{P}_\text{BaTiO$_3$}$ and, consequently, to $\sigma_\text{inter}$. The two possible orientations of the spontaneous polarization thereby lead to polar discontinuities with different magnitudes; see Fig.~\ref{fig:hp_vs_uhp}b and c for the SmO–TiO$_2$ interface. 

If the spontaneous polarization points away from the SmO–TiO$_2$ interface (or towards the NiO$_2$–BaO interface), the interface charges are nearly compensated, given that the spontaneous polarization of BaTiO$_3$ strained to SrTiO$_3$ is $\SI{44.6}{\micro\coulomb\per\centi\meter\squared}$ \cite{simmen_interplay_2025}. This configuration, referred to as the `happy' orientation in the literature  \cite{efe_happiness_2021}, is therefore energetically favored and leads to a vanishing critical thickness for the spontaneous polarization of BaTiO$_3$ on SmNiO$_3$ \cite{simmen_interplay_2025, efe_polarizing_2026}.

In contrast, the opposite `unhappy' orientation is energetically highly disfavored \cite{simmen_interplay_2025, spaldin_layer_2021}. If the polarization points towards the SmO–TiO$_2$ interface, as illustrated in Fig.~\ref{fig:hp_vs_uhp}c, (or away from the NiO$_2$–BaO interface) the interface charges nearly double to $\sigma_\text{inter} \simeq \SI{100}{\micro\coulomb\per\centi\meter\squared}$. In this configuration, the surface charge from the spontaneous polarization approximately doubles the original polar discontinuity. 

\begin{table}[]
    \centering
    \setlength{\tabcolsep}{7pt}
    \begin{tabular}{cccc}
        \toprule
        Material    & Formal charge & Layer charge      & $P_\text{bulk}$ \\ 
                    & A/B           & ($e$)             & ($\SI{}{\micro\coulomb\per\centi\meter\squared}$)      \\ \midrule
        SmNiO$_3$   & +3/+3         & $+1$ (SmO)        & $\sim\SI{50}{} + n 100 $    \\ 
                    &               & $-1$ (NiO$_2$)    &                           \\
        BiNiO$_3$   & +4/+2         & $+2$ (BiO)        & $\sim\SI{0}{} + n 100$        \\
                    &               & $-2$ (NiO$_2$)    &              \\ \bottomrule
    \end{tabular}
    \caption{Summary of the SmNiO$_3$ and BiNiO$_3$ bulk properties. The formal charge of the charge disproportionated site is given as the average value. This average value is also used to calculate the layer charge and the bulk polarization. The polarization lattice is given for a 5-atom unit cell with $a=\SI{4}{\AA}$. }
    \label{tab:overview_bulk}
\end{table}

\paragraph{Screening by changing the disproportionation pattern–}
The unhappy orientation is highly unstable, requiring twice the charge (\SI{1}{\elementarycharge\per\text{f.u.}}) to screen the \SI{\sim 100}{\micro\coulomb\per\centi\meter\squared} polar discontinuity. In our DFT calculations
of the SmNiO$_3$/BaTiO$_3$ interface, we find a modification of the charge disproportionation pattern of SmNiO$_3$ as an alternative, low-energy screening mechanism to the usual formation of a 2DEG or defects. 

We perform DFT calculations on a superlattice constructed from 5 layers of BaTiO$_3$ and 3 bilayers of SmNiO$_3$ out-of plane, with $\sqrt{2}\times\sqrt{2}$ formula units in plane (110 atom supercell). We initialize the spontaneous polarization of BaTiO$_3$ in either the up (happy) or down (unhappy) direction. Given that the unhappy orientation is energetically unfavorable, we constrain the polarization of the middle three BaTiO$_3$ layers to the bulk polarization value by fixing the corresponding atoms during the relaxation. 

\begin{figure}
    \centering
    \includegraphics[width=\linewidth]{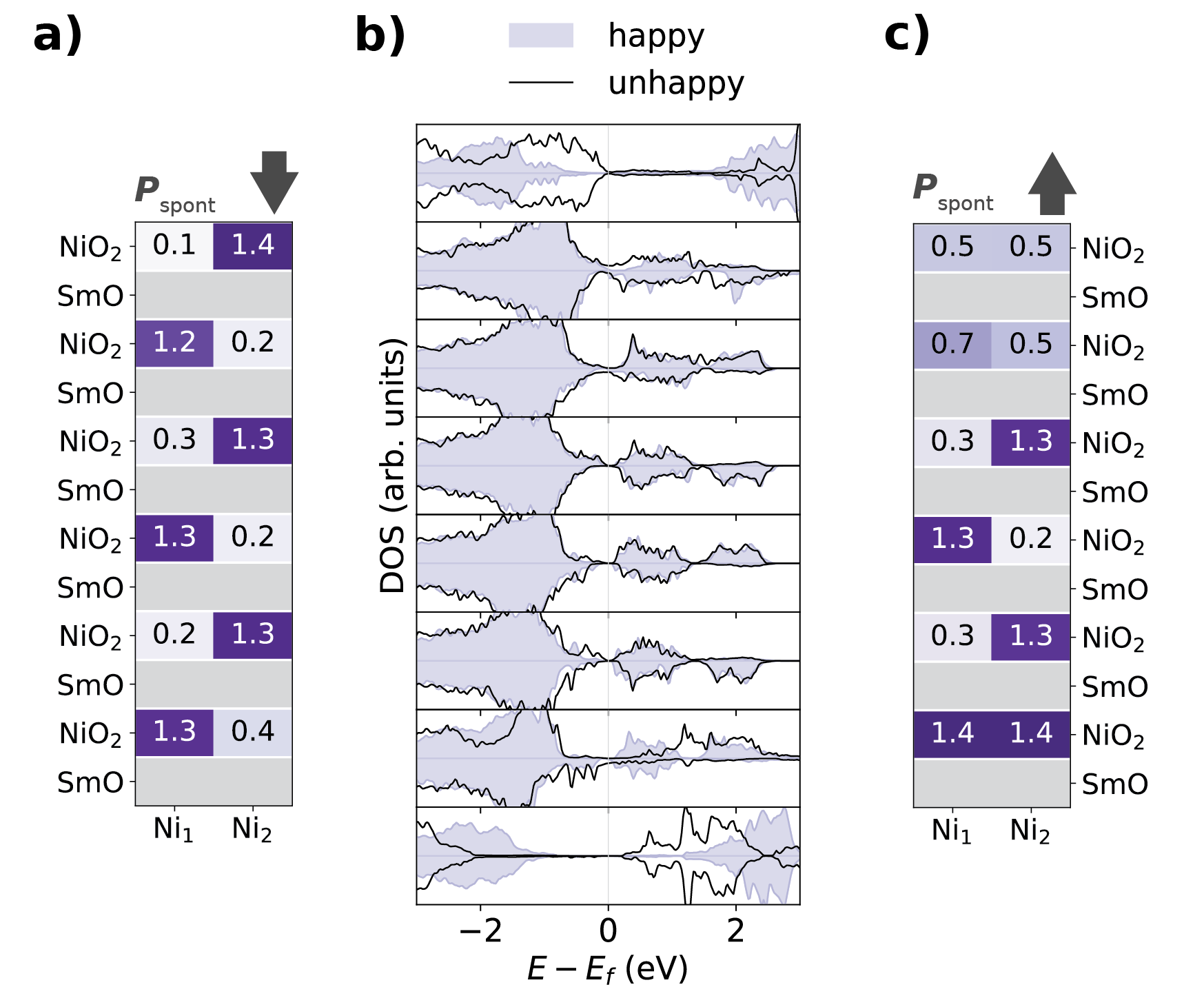}
    \caption{Ni-$d$ magnetic moments in the happy (a) and unhappy (b) orientation. The disproportionation pattern in the unhappy orientation is changed to formal charges of Ni$^{4+}$/Ni$^{4+}$ at the upper NiO$_2$–BaO interface and Ni$^{2+}$/Ni$^{2+}$ at the lower SmO–TiO$_2$ interface. b) Comparison of the DOS of the happy (purple area) and unhappy (black line) orientation. Only the two interfacing layers of the BaTiO$_3$ slab are shown (layers 1 and 8).  }
    \label{fig:sno_bto_magmom_dft}
\end{figure}

When the polarization is in the unhappy orientation, we observe a modification of the charge disproportionation at the interface. 
Fig.~\ref{fig:sno_bto_magmom_dft}  shows the magnetic moments on the Ni atoms in the SmNiO$_3$ slab of the superlattice, serving as a proxy for the Ni formal charge. The bulk checkerboard disproportionation pattern is clearly present in the bulk region of the SmNiO$_3$ slab for both polarization directions. In the happy orientation (Fig.~\ref{fig:sno_bto_magmom_dft}a), this checkerboard pattern is preserved up to the interface. However, when the spontaneous polarization points in the opposite, unhappy orientation (Fig.~\ref{fig:sno_bto_magmom_dft}c) the checkerboard disproportionation pattern is disrupted; with the two Ni sites closest to each interface becoming equivalent to each other. At the upper NiO$_2$–BaO interface, the magnetic moments approach zero (two Ni$^{4+}$ sites) while the magnetic moments are close to $\SI{1.4}{\mu_B}$ for both Ni sites at the lower SmO–TiO$_2$ interface (two Ni$^{2+}$ sites).

\begin{figure}
    \centering
    \includegraphics[width=\linewidth]{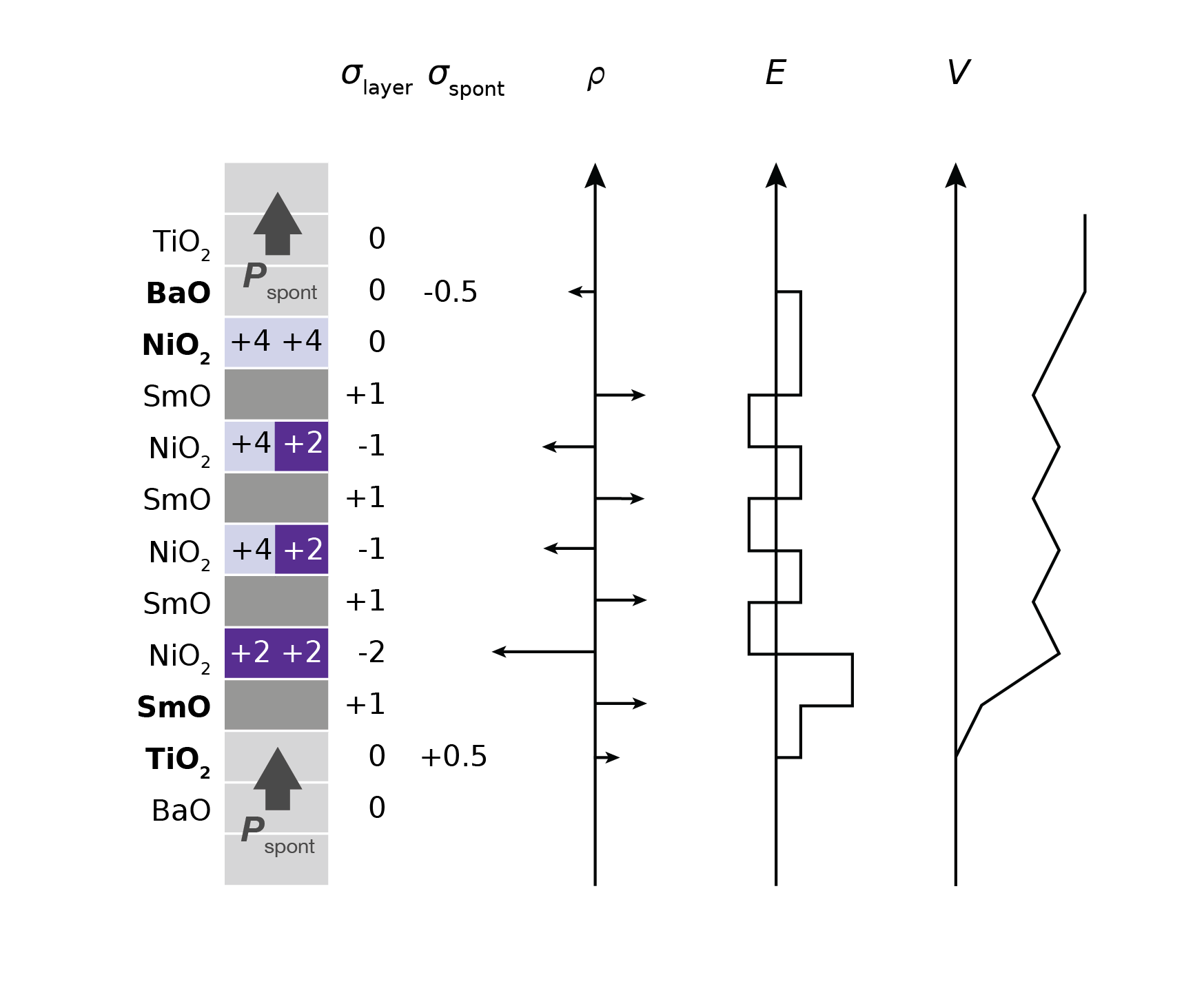}
    \caption{Changing the charge disproportionation pattern to +4/+4 and +2/+2 in the NiO$_2$ sublayers adjacent to the respective NiO$_2$–BaO (top) SmO–TiO$_2$ interface (bottom) screens the polar discontinuity in the unhappy orientation. The new charge density ($\rho$) leads to electric fields ($E$) centered around zero and a non-diverging potential ($V$). }
    \label{fig:change_disprop_pattern_theory}
\end{figure}

This modified charge disproportionation pattern formally corresponds to the transfer of \SI{1}{\elementarycharge\per\text{f.u.}} between the two interfaces and can screen the increased polar discontinuity in the unhappy orientation, as illustrated in Fig.~\ref{fig:change_disprop_pattern_theory}. Transferring \SI{2}{\elementarycharge} to the NiO$_2$ sublayer closest to the lower SmO–TiO$_2$ interface doubles its negative layer charge to $-2$, centering the electric field around zero and keeping the potential around a constant value. Removing \SI{2}{\elementarycharge} from the NiO$_2$ sublayer adjacent to the upper NiO$_2$–BaO interface effectively inverts the sign of the polar discontinuity and shifts its location one sublayer deeper into the SmNiO$_3$ slab, thus compensating the interface charge. Note that, despite being screened, we expect the unhappy polarization orientation to remain energetically disfavored due to the energy cost associated with the modified disproportionation pattern.

Next to electrostatics, the formal charges of interfacial ions are also affected by the chemical bonding across the interface. We find that the charge disproportionation is disrupted in more than one NiO$_2$ sublayer next to the NiO$_2$–BaO interface, and the Ni magnetic moments in these sublayers ($\sim \SI{0.5}{\mu_B}$ at the NiO$_2$–BaO interface) are larger than in the bulk part of the slab ($\sim\SIrange{0.2}{0.3}{\mu_B}$), which cannot be explained by electrostatics. However, we cannot determine the exact amount of transferred charge and differentiate between these competing effects due to hybridization of nickel with the surrounding oxygen.

The layer-projected DOS are consistent with the results from the magnetic moments; see Fig.~\ref{fig:sno_bto_magmom_dft}b. Charge disproportionation is present in the bulk-like region (layers 2-4) regardless of the polarization orientation, leading to a visible band gap and spin-polarized states near $\SI{2}{eV}$. In the interface region, the disproportionation pattern is preserved in the happy orientation (purple area) and disrupted in the unhappy orientation (black line). At both interfaces, we see changes in the DOS that cannot be explained by the modified charge disproportionation pattern. At the upper NiO$_2$–BaO interface, some spin-polarized states are retained around \SI{2}{eV}, consistent with a lower formal charge and the larger Ni magnetic moments seen previously. Since this NiO$_2$ sublayer is directly bonded to the BaTiO$_3$, this behavior is likely due to a difference in chemical environment between the two opposite polarization directions. 

The modified charge disproportionation pattern increases the width of the visible interface region in the DOS and the magnetic moments. For example, we find a band gap in the DOS of the happy orientation in all layers except the one adjacent to the NiO$_2$–BaO interface. In the unhappy orientation, the band gap disappears in 1-2 layers next to both interfaces. We can explain this behavior with our simple electrostatic model in Fig.~\ref{fig:change_disprop_pattern_theory}: The change in disproportionation pattern distributes the interface charge over a longer distance before screening is achieved, increasing the interface width. 

\begin{figure}
    \centering
    \includegraphics[width=\linewidth]{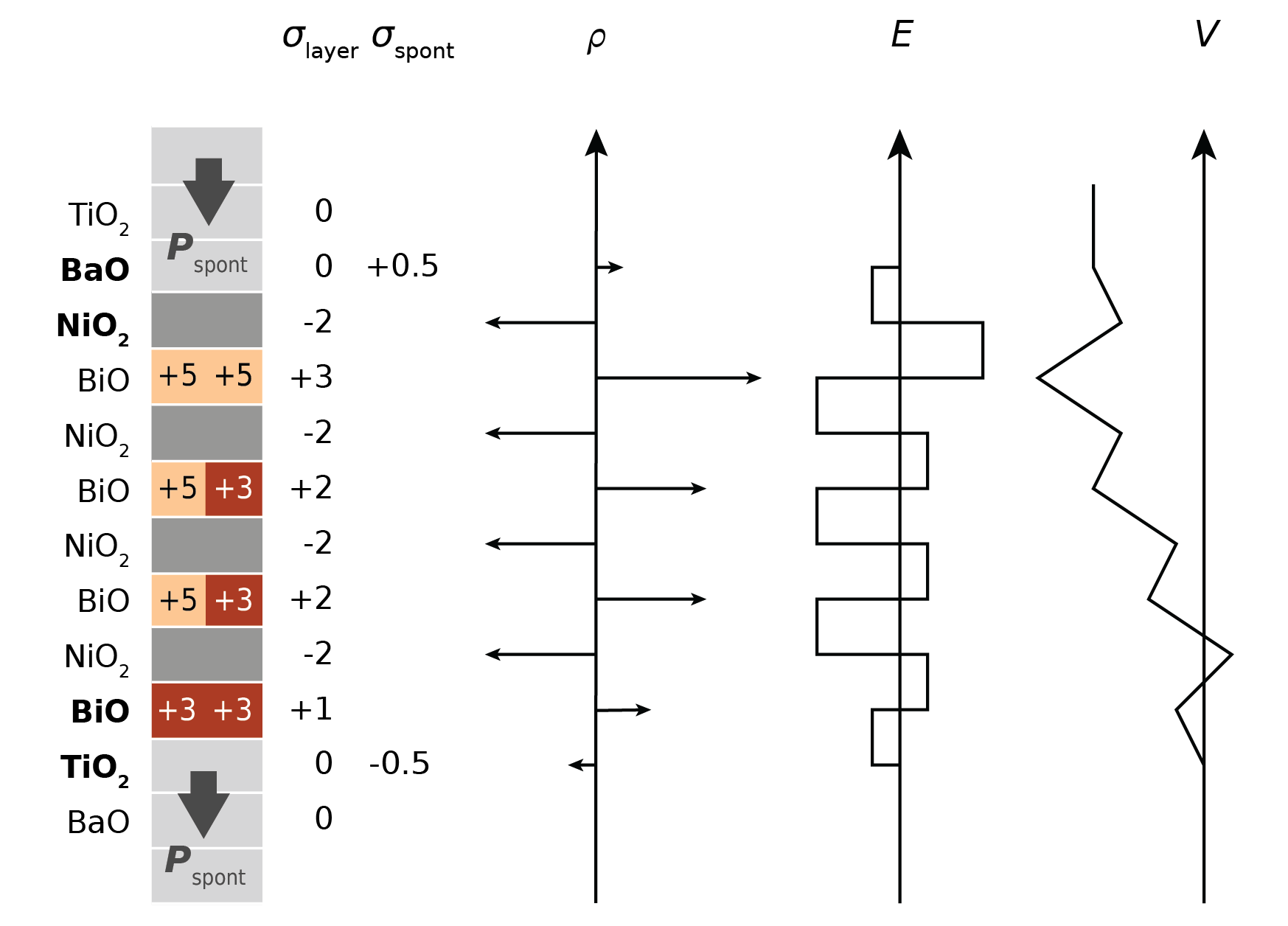}
    \caption{The polar discontinuity in BiNiO$_3$/BaTiO$_3$ superlattices cannot be compensated by the spontaneous polarization. Additional screening by changing the charge disproportionation in the happy orientation overcompensates the polar discontinuity, leading to a diverging potential ($V$).
    }
    \label{fig:bno_bto_with_ct}
\end{figure}

Analogous screening mechanisms could also exist in other charge disproportionated materials, such as other members of the rare-earth nickelate family or perovskites such as CaFeO$_3$ \cite{woodward_structural_2000}, BaBiO$_3$ \cite{lobo_bismuth_1995}, or BiNiO$_3$ \cite{ishiwata_pressureinduced_2004, azuma_pressureinduced_2007}. However, different changes to the charge disproportionation pattern might be needed for full screening of the polar discontinuity. In BiNiO$_3$, for example, the Ni ion is formally +2 charged and the charge-disproportionated Bi (Bi$^{3+}$ and Bi$^{5+}$) has an average formal charge of +4 \cite{ishiwata_pressureinduced_2004, azuma_pressureinduced_2007}, resulting in formal (001) layer charges of $\pm\SI{2}{\elementarycharge}$ and a polar discontinuity of $\sigma_\text{inter} \simeq \pm \SI{100}{\micro\coulomb\per\centi\meter\squared}$ when interfaced with nonpolar BaTiO$_3$. Since the polar discontinuity is twice as big as in BaTiO$_3$/SmNiO$_3$, the spontaneous polarization of BaTiO$_3$ provides only half the compensation charge needed for screening the polar discontinuity with BiNiO$_3$.  

For BiNiO$_3$, changing the charge disproportionation by transferring one electron per formula unit from one interface to the other would either overcompensate (happy orientation) or not fully compensate (unhappy orientation) the polar discontinuity between BiNiO$_3$ and BaTiO$_3$. \SI{1}{\elementarycharge\per\text{f.u.}} is twice the charge needed to compensate the polar discontinuity of $\pm \SI{50}{\micro\coulomb\per\centi\meter\squared}$ in the happy orientation, as illustrated in Fig. \ref{fig:bno_bto_with_ct}a. In the opposite, unhappy orientation, the spontaneous polarization increases the polar discontinuity to $\pm \SI{150}{\micro\coulomb\per\centi\meter\squared}$ and a transfer of \SI{1}{\elementarycharge\per\text{f.u.}} is too small for full compensation of the interface charges. For this system, full screening of the polar discontinuity could only be achieved with either a larger spontaneous polarization ferroelectric such as supertetragonal BaTiO$_3$ \cite{mellaerts_origin_2022}, or from a more complex charge disproportionation pattern or a larger spontaneous polarization. For example, in the happy orientation, 1:3 and 3:1 ratios of Bi$^{3+}$:Bi$^{5+}$ at the BiO-TiO$_2$ and NiO$_2$–BaO interfaces, respectively, would formally compensate the interface charge, but might introduce unfavorable lattice strains. 

In summary, we have demonstrated a novel screening mechanism for the polar discontinuity between insulating SmNiO$_3$ and ferroelectric BaTiO$_3$. In the unhappy orientation, the spontaneous polarization maximizes the polar discontinuity and SmNiO$_3$ screens by adjusting its charge disproportionation pattern from +2/+4 to +2/+2 and +4/+4 at the SmO–TiO$_2$ and the NiO$_2$–BaO interface, respectively. Although we found this to be the predominant screening mechanism, changes in the chemical environment also affected the disproportionation of Ni, especially at the NiO$_2$ -BaO interface. We briefly discussed the polar discontinuity between BaTiO$_3$ and the $A$-site disproportionated BiNiO$_3$, highlighting that changing the charge disproportionation pattern might not always be sufficient to screen the polar discontinuity. Our results highlight the potential of interface charge engineering to stabilize unconventional charge-disproportionation patterns and metastable oxidation states. \\ 

\textit{Acknowledgements--} This work was supported by the Swiss National Science Foundation under Grant No.~209454 and by ETH Zurich. The calculations for this work were performed on the Swiss National Supercomputing Center Daint cluster under Project No.~lp61. \\

\textit{Data availability--}
The relevant input files and data of our ab initio calculations are openly available on the Materials Cloud Archive at \url{http://doi.org/DOItobeinsertedhere}.

\bibliography{references}

\newpage

\section{Supporting information }

\subsection{Methods}

We performed density-functional theory calculations with the VASP code \cite{kresse_efficiency_1996, kresse_efficient_1996}, using the Perdew-Burke-Ernzerhof functional revised for solids (PBEsol) \cite{perdew_restoring_2008} to describe the exchange correlations. We treated 10 electrons for Ba (5$s^2$5$p^6$6$s^2$), 12 for Ti (3$p^6$4$s^2$3$d^4$), 6 for O (2$s^2$2$p^4$), 16 for Ni (3$p^6$4$s^2$3$d^8$) and 11 for Sm (5$s^2$5$p^6$6$s^2$5$d^1$) as valence electrons in the Projector augmented-wave (PAW) pseudopotentials \cite{blochl_projector_1994, kresse_ultrasoft_1999}, with the $f$ electrons of Sm frozen into the core. We ran the calculations using the double-relaxation, static and band-structure workflows for VASP from atomate2 \cite{ganose_atomate2_2025}.

The bulk structures were relaxed in a 20-atom unit cell until the forces on all the atoms were smaller than \SI{0.01}{\electronvolt\per\angstrom}, with the in-plane lattice-parameters of SmNiO$_3$constrained to to those of a SrTiO$_3$ substrate (cubic lattice constant $a_{pc} = \SI{3.898}{ \AA}$ \cite{wahl_srtio3_2008}). A $\Gamma$-containing $k$-point grid was employed in all bulk calculations, with a mesh of $8\times8\times 6$ $k$-points for relaxations and $12\times12\times10$ $k$-points for DOS calculations.  A correction according to the formulation by Liechtenstein \textit{et al.} \cite{liechtenstein_densityfunctional_1995} was applied on the Ni-$d$ states. A good agreement of the charge disproportionation in SmNiO$_3$ with experimental data was achieved for $U = \SI{2}{\electronvolt}$ and $J = \SI{1}{eV}$, as previously reported in literature \cite{hampel_interplay_2017}. We approximated the real magnetic order of SmNiO$_3$ with $A$-type antiferromagnetic order (AFM), since the $S$ and $T$-type AFM magnetic orders compatible with experiments \cite{giovannetti_multiferroicity_2009, hampel_interplay_2017} are too large for applications in a supercell. The $A$-AFM ordering was shown to give very similar results compared to more complicated AFM orders \cite{hampel_interplay_2017} and is the simplest AFM ordering where the $P2_1/c$ phase is still stable  \cite{varignon_complete_2017}. 

All superlattices consist of 110 atoms (unless otherwise stated), with $\sqrt{2}\times\sqrt{2}$ formula-units in-plane, 3 unit cells of SmNiO$_3$ out-of plane (corresponding to 6 layers or formula units out of plane) and 4 unit cells of BaTiO$_3$. We used the same calculation parameters as for the bulk structures, but included only 1 $k$-point along the $c$ axis. For DOS calculations, the in-plane $k$-points were doubled to $16\times16\times1$. The superlattices were strained to the same in-plane lattice parameters as the bulk, resulting in around \SI{-2.2}{\%} compressive and \SI{3.8}{\%} tensile strain for bulk \ce{BaTiO3} and \ce{SmNiO3} unit cells constrained to SrTiO$_3$ compared to the relaxed bulk unit cells.

\end{document}